\documentclass{aastex631}

\begin{document}
%Key effort: add more citations! 
%--------------------------------------------------------
\title{When Magnetic Field Lines Stretch, Snap, and Expand: A New Look at Solar Flares with L-maps}
%-
%-Title 2: From Slow Rise to Eruption: Three Stages of  Solar-Flare Coronal Magnetic Evolution Revealed by L-maps; 
%Title 3: Tracking the 3D Magnetic Evolution of Solar Flares Using L-maps from Data-Driven Simulations
%\title{From Slow Rise to Eruption: Three Stages of  Solar-Flare Coronal Magnetic Evolution Revealed by L-maps}
%\title{Tracking the 3D Magnetic Evolution of Solar Flares Using L-maps from Data-driven Simulations}
%\title{Evolution of Flare Ribbons and Coronal Dimmings During an X-class Flare Using L-maps in Realistic MHD Simulations and Observations}
%-----------------------------------------------------------
\author[0000-0001-8975-7605]
{Maria D. Kazachenko}
\affiliation{Laboratory for Atmospheric and Space Physics, University of Colorado Boulder, Boulder, CO 80303, USA}
\affiliation{National Solar Observatory, 3665 Discovery Drive, Boulder, CO 80303, USA}
\affiliation{Department of Astrophysical and Planetary Sciences, University of Colorado Boulder, 2000 Colorado Avenue, Boulder, CO 80305, USA}

\author[0000-0001-8975-7605]{Yuhong Fan}
\affiliation{High Altitude Observatory, NSF National Center for Atmospheric Research, 3080 Center Green Drive, Boulder, CO 80301, USA}

\author[0000-0002-9667-6907]{Andrey N. Afanasyev}
\affiliation{Laboratory for Atmospheric and Space Physics, University of Colorado Boulder, Boulder, CO 80303, USA}
\affiliation{National Solar Observatory, 3665 Discovery Drive, Boulder, CO 80303, USA}
\affiliation{Institute of Solar-Terrestrial Physics of SB RAS, Irkutsk, Russia}
%-----------------
\begin{abstract}
%Understanding the three-dimensional evolution of coronal magnetic fields during solar flares is challenging due to the lack of direct coronal field measurements. In this study, we combine data-driven MHD simulations of NOAA AR 11158 \citep{Fan2024} with flare ribbon and coronal dimming observations to characterize realistic coronal magnetic field evolution during an X-class flare. We introduce \textit{L-maps}, maps of the logarithm of field-line lengths, to track simulated coronal magnetic structures. L-maps allow identification of flare ribbons (shortening field lines) and coronal dimmings (lengthening field lines). Comparison with SDO/AIA observations validates the simulation. Using K-means clustering of L-map temporal profiles, we identify three stages of coronal magnetic evolution: slow pre-flare rise phase, flare reconnection with CME rise, and post-reconnection CME expansion. We find slow pre-flare rise phase of ribbon field lines and discover a new phenomenon, a rapid expansion of reconnected inner flux-rope field lines, which we term ``reconnection dimmings.'' We conclude that L-maps allow us to connects realistic simulations with observations and provide a robust method to track 3D coronal magnetic field evolution.
Understanding the three-dimensional evolution of coronal magnetic fields during solar flares remains challenging due to the lack of direct coronal field measurements. Here we combine data-driven MHD simulations of NOAA AR~11158 \citep{Fan2024} with flare-ribbon and coronal-dimming observations to investigate realistic coronal magnetic-field evolution during an X-class flare. We introduce {\it L-maps} — { maps of natural logarithm of magnetic field-line lengths} — as a diagnostic tool to track the dynamics of simulated coronal magnetic structures. Variations in L-maps identify flare ribbons through field-line shortening and coronal dimmings through field-line lengthening. Comparison with SDO/AIA observations demonstrates strong morphological and temporal agreement, validating the simulated field evolution. Applying K-means clustering to the L-map temporal profiles, we distinguish three stages of coronal evolution: (1) slow pre-flare rise phase, (2) flare reconnection accompanied by CME rise, and (3) post-reconnection CME expansion. We detect a slow pre-flare rise phase of magnetic field lines routed in ribbon footpoints and identify {\it reconnection dimming}—{ area of rapid expansion of active-region core magnetic field lines during flare impulsive phase due to reconnection}. Our results show that L-maps provide a powerful and physically intuitive framework for bridging simulations and observations and for tracking the full three-dimensional evolution of coronal magnetic fields during flares.
\end{abstract}


\section{Introduction} \label{sec:intro}
%Broad context
Stellar flares and stellar mass ejections \citep{Kowalski2024} lead to energetic events that affect surrounding exoplanets and their habitability \citep{Segura_2018_haex_book,Shields2019,Cliver+2022}. Within our solar system, these processes can be studied in unparalleled detail thanks to the direct observations of solar coronal mass ejections (CMEs, e.g., \citealt{Gosling1974}) and flares, which expel large quantities of magnetized plasma ($\sim$$10^{11}$–$10^{13}$~kg) along with bursts of electromagnetic radiation and energetic particles into the heliosphere \citep{Forbes2006,Benz2008}. When directed toward Earth, CMEs can strongly interact with the magnetosphere and ionosphere, disrupting satellite operations and communications, inducing currents that overload power grids, and posing radiation hazards to astronauts \citep{Temmer2021}.

%{\it Key Questions of Flare Physics:} 
Despite decades of solar observations, the detailed three-dimensional dynamics of magnetic fields during flares and CMEs, including their triggering and the formation of erupting structures, remains unclear \citep{Janvier2015,Wang2015review,Toriumi2019,Kazachenko2022r,Georgoulis2024}.  A major obstacle is the lack of direct magnetic field measurements in solar corona. { One possibility to constrain three-dimensional dynamics of magnetic fields during flares is using}  simulations to understand flare/CME onset and evolution. 
The Solar Dynamics Observatory (SDO, \citealt{Pesnell2012}) enabled a major advance by providing the first routine vector magnetic field measurements in the photosphere, allowing time-dependent modeling of the coronal magnetic field. Among the most realistic of these models are data-driven simulations that use observed electric, velocity, or magnetic fields in the photosphere as lower-boundary  conditions. These quantities can be derived from time sequences of vector magnetic field and Doppler velocity measurements using electric- or velocity-field inversion techniques (e.g.,  \citealt{Schuck2008,Lumme2019,Fisher2020,Afanasyev2021}).
Since the first SDO-based data-driven simulation by \citet{Cheung2012}, this approach has become one of the most realistic approaches to advance our understanding of flare magnetism \citep[e.g.,][]{Hoeksema2020,Lumme2022,Afanasyev2023,Fan2024}, as summarized in recent reviews \citep{Jiang2022,Schmieder2024}. 

%{\it Validating 3D Coronal Magnetic Field Simulations:} 
As the realism of 3D MHD models has increased, so has the need for quantitative tools to compare and validate simulations against observations.
%However, as the realism of 3D MHD data-driven simulations has increased, it became evident that coronal magnetic fields have complex structure and one needs good metrics to compare and validate these simulations using observations in both qualitative and quantitative ways. 
%
%In the following, we summarize some of the most widely used comparison/validation approaches.   
The majority of existing validation efforts have focused on qualitative comparisons between simulated coronal magnetic structures or some kind of emission proxies and loops observed in extreme ultraviolet by, e.g., the Atmospheric Imaging Assembly (AIA) at Solar Dynamics Observatory (SDO). For example, \citet{Cheung2012} proposed using line-of-sight integrated current density as a synthetic emissivity proxy. \citet{Fan2024} and many others employed forward modeling to generate synthetic AIA images in different coronal channels. Alternatively, since coronal magnetic fields cannot be measured directly, it is valuable to use indirect observational proxies, such as chromospheric ribbons and coronal dimmings, that trace the evolution of magnetic connectivity during eruptions. For example, multiple works looked at proxies for ribbons dynamics, using evolution temperature maps \citep{Afanasyev2023}, quasi-separatrix-layers \citep{Fu2023,Guo2023,Guo2024} and current density maps \citep{Liu2025} at the lower boundary of simulation domain and compared these with ribbons' observations to validate the model. These diverse approaches highlight the growing effort to bridge simulations and observations.

Building on these efforts, flare ribbons and coronal dimmings provide particularly powerful proxies, as they trace the footpoints of reconnected magnetic field lines and expanding coronal structures, respectively, offering unique insights into flare and CME feedback that are otherwise inaccessible \citep{Dudik2025}.
Ribbons could be observed as enhanced emission in, e.g., H$\alpha$ and 1600 \AA{} channels in the transition region and the upper chromosphere, marking the response to energy deposition at $\sim$2000~km heights \citep{Cheng1983,Doschek1983,Forbes1991_rcc}. { Dimmings, could be observed as dim areas in e.g., 193~\AA{}, 211~\AA{} and 304~\AA{} channels primarily in the solar corona, marking the significant depletion in coronal EUV emission due to plasma evacuation along expanding field lines} (see e.g., \citealt{Mason2019,Dissauer2018,Krista2017} and recent review  on dimmings by \citealt{Veronig2025}).  

{ 
In particular, intensity evolution within coronal-dimmings could be used to infer the evolution of overlying magnetic structures. For example, \citet{Cheng2016,Aschwanden2017a} used time-dependent EUV dimmings to estimate expansion of erupting coronal structures during the flare, while \citet{Wang2019} tracked pre-flare dimming across multiple wavelengths to infer the stretching and opening of coronal fields before the flare. Separately, using SDO/AIA 304\AA{} observations \citet{Qiu2024} analyzed the tempo-spatial sequences of both brightenings and dimmings in the upper chromosphere and transition region, to track dynamic evolution of magnetic fields before and during eruptive flares, such as expansion and reconnection. %They found that some dimmings are preceded by impulsive brightenings, indicating that fields lines in these regions first reconnect and then rapidly expand.

In this paper, we introduce maps of natural logarithm of magnetic field-line lengths, or {\it L-maps}, applied to the output of 3D data-driven MHD simulations to show how these, together with flare ribbons and dimming observations, could be used to understand complex evolution of 3D coronal magnetic fields of solar eruptions.}
%
%As proxies of magnetic field line strength, both ribbons and dimmings could be identified in the simulations and compared with simulations. %Furthermore, the evolution of magnetic flux swept by flare ribbons and dimmings could be used to track reconnection flux (and its rate) and magnetic flux of an expanding flux rope, respectively, in a quantitative way.
%
%{\it Relationship between Ribbons/Dimmings and CME properties:}  Many statistical studies have found correlations between cumulative ribbon, dimming, and CME properties indicating that magnetic reconnection is key in the CME acceleration process, especially for fast CMEs (see review paper by \citealt{Kazachenko2022r}). Furthermore, temporal profiles of the CME liftoff and reconnection and dimming fluxes suggest that both the ideal MHD instability and reconnection play major roles in the CME acceleration. In addition, two properties, the reconnected fraction of the AR flux and the total AR magnetic flux, were shown to be the two key parameters defining whether a flare would be confined or eruptive \citep{Li2022,Kazachenko2023}. 
%\citep{Kahler2017,Kahler2017}. 
%
%PAPER STRUCTURE
Our paper is structured as follows.
%mhd & obs
In Sections~\ref{sec:sim1} and Section~\ref{sec:obs} we describe the methods applied to simulations and observations, respectively.
%results
In Section~\ref{sec:results} we describe the results.
%discussion
In Sections~\ref{sec:discussion} and \ref{sec:conclusion} we discuss our results and draw conclusions.

\section{Data \& Methods}
In this paper, we analyze an X2.2 flare that occurred in NOAA AR 11158 on February 15, 2011 using data-driven MHD simulations and observations. According to GOES the flare started at 01:44 UT, peaked at 01:56 UT, and ended at 02:06 UT and was accompanied by a front-side halo coronal mass ejection (CME). In Section \ref{sec:sim1}, we describe the setup and parameters of the data-driven simulations, and in Section \ref{sec:obs} the details of the observational data and processing methods used to identify ribbons and dimmings.

\subsection{Data-Driven MHD Simulations}\label{sec:sim1}
For our analysis of coronal magnetic field evolution, we use output from the data-driven MHD simulation of the 2011 February 15 X2.2 flare and associated CME in active region (AR) NOAA 11158, as described by \citet{Fan2024} and \citet{Linton2023}. 
The simulation is carried out in a spherical wedge domain with a longitudinal/latitudinal widths of $18.8^\circ\times17.2^\circ$, and a radial range of $1.0~R_\odot$ to $1.43~R_\odot$.
The simulation is driven at the lower boundary with an electric field derived from the observed normal magnetic field \citep{Kazachenko2014,Kazachenko2015,Fisher2020} and the vertical electric current density, $J_z$, derived from the horizontal magnetic field from the Helioseismic and Magnetic Imager (HMI, \citealt{Scherrer2012}) onboard the SDO. This approach ensures that the driving boundary conditions reproduce the observed photospheric evolution of electric currents and magnetic shear.

The model captures the build-up of a pre-eruption coronal magnetic field configuration that closely resembles a nonlinear force-free field extrapolation and produces multiple eruptive events over the simulation period. The pre-flare coronal field consists of highly sheared and twisted magnetic field lines above the polarity-inversion line (PIL), showing strong qualitative agreement with the hot coronal loops observed in the SDO/AIA 94~\AA{} and 131~\AA{} channels.

\begin{figure*}[!htb]
% Answer: [trim={left bottom right top},clip]
\centering 
\includegraphics[width=18.0cm,trim={0.0cm 0 0.0cm 0.0cm},clip]{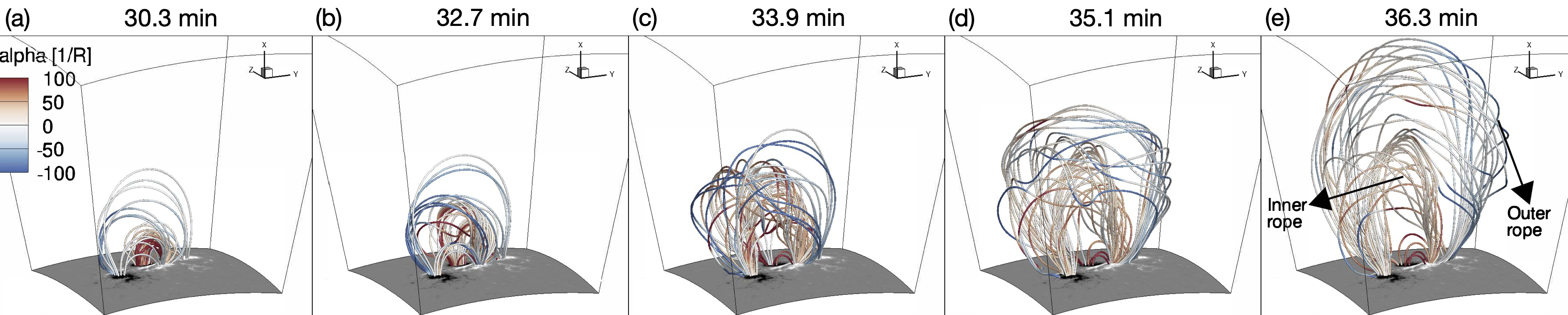}
\caption{Evolution of the three-dimensional coronal magnetic field during the simulated eruption of an X2.2 flare in NOAA~AR~11158 on 2011~February~15. The sequence illustrates the slow pre-flare rise phase (panel~a), flare reconnection and CME onset (panels~b--c), and post-reconnection CME expansion (panels~d--e). Field lines are traced from Lagrangian tracer points advected by the velocity field and are color-coded by their twist rate. The grayscale background shows the normal component of the SDO/HMI magnetogram. { See \S~\ref{sec:sim1} for details.} An animated version of this figure is available in \citet{Fan2024}. }
\label{fig:3dcorona}
\end{figure*}

Figure~\ref{fig:3dcorona} illustrates the evolution of the simulated three-dimensional coronal magnetic field evolution during eruption \citep{Fan2024}. We find that the eruption is initiated by tether-cutting reconnection within the sheared core field above the PIL, leading to the formation of a magnetic flux rope (FR) with dipped field lines that subsequently erupts. The resulting structure consists of two oppositely twisted flux ropes, producing an outgoing double-shell feature in EUV images consistent with that seen in the STEREO-B/EUVI 171 observations of the CME. The footpoints of the erupting field lines spatially correspond to the coronal dimming regions seen in the SDO/AIA 211~\AA{} observations { \citep{Schrijver2011}}.

The close correspondence between the simulated and observed magnetic morphology and dimming evolution demonstrates that the electric-field driving approach, constrained by the measured $J_z$, provides a physically consistent and promising method for reproducing realistic coronal magnetic-field evolution during solar eruptions.

%In Figure~\ref{fig:3dcorona} we show the evolution of the 3D coronal magnetic field during the eruption from \citet{Fan2024}.  We find that the eruption is triggered by the tether-cutting reconnection in a highly sheared field above the polarity-inversion line (PIL) and a magnetic flux rope with dipped field lines forming during the eruption. The final modeled erupting structure contains two oppositely directed flux ropes that result in a double shell white-light image consistent with STEREO observations of the CME.  The foot points of the erupting field lines correspond to the dimming regions seen in the SDO/AIA 211 \AA{} channel. These agreements of simulations with observations suggest that the selected electric-field driving approach matching $J_z$ is a promising way to drive realistic coronal magnetic field evolution during eruptions. 

%Motivation statement TBD
To identify the locations of flare ribbons and coronal dimmings and to track their temporal evolution in the simulations, we developed and implemented an {\it L-map approach}. The L-map, $L(x,y,t)$, is defined as the natural logarithm of the magnetic field line length, 
$l(x,y,t)$ \citep{Lynch2021, Kazachenko2022, Dahlin2025}:
\begin{equation}
L(t) \equiv L[x, y, t] = \ln{[l(x, y, t)]},
\end{equation}
{ where the field line length $l$ is normalized to units of solar radius $R_\odot$}.

To calculate these field line lengths, we traced field lines at $3$ Mm above the lower boundary of the simulation domain, which  corresponds to the transition to the corona. The key advantage of using L-maps is that they allow us to quantify temporal changes in magnetic field-line connectivity:
\begin{equation}
\Delta L(t) = L(t) - L(t - \Delta t)=\ln\left(\frac{l(t)}{l(t-\Delta t)}\right),
\end{equation}
which capture both abrupt reconfigurations associated with magnetic reconnection and more gradual field expansion or implosion \citep{Hudson2008,Yadav_Kazachenko_2023_ApJ}.

We use the temporal running difference between consecutive  L-maps to track ribbon and dimming signatures in the simulations.
To identify { flare-ribbon proxies}, we locate regions where magnetic field line lengths decrease sharply over time, corresponding to magnetic reconnection and subsequent field-line shortening by more than a defined threshold, $\Delta L_\mathrm{thr,ribbons}$, \citep{Lynch2021, Kazachenko2022}.
To identify { coronal-dimming proxies}, we select regions where the magnetic field line lengths increase by more than a defined threshold, $\Delta L_\mathrm{thr,dimming}$, indicating field expansion during or after the eruption.

%We use L-maps to track ribbons and dimmings evolution in the simulations as described below.
%{\it Ribbon proxy (simulations):} 
%To find ribbon proxies, we identify areas where field lines decrease their length (due to reconnection and subsequent shortening, \citealt{Lynch2021,Kazachenko2022}). 

%UNCERTAINTIES
%We vary time step $\Delta t$ and length threshold $\Delta L_\mathrm{thr,rbn}=0.4R_\odot$ to identify the uncertainty in the ribbon proxy.  
%As a test to using L-maps to identify ribbons, we also identify ribbons using temperature maps at the base of the corona, i.e. 3 Mm above the lower boundary \citep{Afanasyev2023}. In this case, we identify ribbon-proxies as locations in the lower boundary of our simulation domain, where temperature is either above a certain temperature threshold, $T_\mathrm{thr}$, or where temperature changes by more than certain temperature $\Delta T_\mathrm{thr}$ within a certain duration $\Delta t$.
%
%{\it Dimming proxy (simulations):}
%To find dimming proxies, we looked for areas where field lines increased their length (due to expansion) by over $\Delta L_\mathrm{thr,dimming}$. 
%In addition, we track response of the magnetic fields at the lower boundary (photosphere) and the base of the corona (3 Mm above) to flare magnetic-field reconfiguration. Specifically, we track changes in the horizontal magnetic field and vertical current density with respect to flare ribbons. 

This L-map approach thus provides a unified quantitative framework for distinguishing and following the spatial and temporal evolution of ribbons and dimmings—key observational signatures of magnetic reconnection and coronal restructuring.

\subsection{Observations: Identifying Flare Ribbons and  Coronal Dimmings}\label{sec:obs}
%{\it Ribbons (Observations):} 

To identify flare-ribbon locations in the observations, we used AIA 1600~\AA{} images and applied the detection { and de-saturation procedures} described by \citet{Kazachenko2017}. At each time step $t_k$, ribbon pixels were defined as those that exceeded a cutoff intensity threshold of $I_c=8$ times the median image intensity.

%{\it Dimmings (Observations):} 
%Uses dim_anal: restore,'../dimmings/aia/aia211_20110215/sav-aia1600/aia1600blos20110215_0130_11158_X2.2_desat.sav',/ver
%dim_anal,faia1600,hmilos,tim,inst_mskarr = inst_mskarr, cum_mskarr = cum_mskarr,dimming_cum_mskarr_run=dimming_cum_mskarr_run,filename='../dimmings/aia/aia211_20110215/sav-aia1600/dim_anal_lin_diff_10_frames.sav',ind1=1,ind2=36,diff_thr = -100.,aia_thr = 200.,log_values=0   

To identify coronal dimming regions from the AIA 211~\AA{} observations, we applied a linear base difference technique to every third frame of the 12~s cadence data, yielding an effective temporal resolution of 36~s \citep{Dissauer2018}. The pre-flare base image was constructed as the median of the first ten images acquired between 01:30~UT and 01:35~UT, several minutes before the onset of the X2.2 flare.
To define dimming pixels, we selected pixels that exhibit both a decrease in intensity relative to the base image that exceeds a threshold of $\Delta I = -100$~DN~pix$^{-1}$ and an absolute intensity below $I_{\mathrm{thr}} = 200$~DN~pix$^{-1}$. These identified pixels were then used to construct instantaneous dimming masks, following the same procedure as for the ribbon mask calculation.

%No fluxes
%Finally, by combining all dimming pixels up to the time $t_k$ we find the cumulative dimming masks. 

\section{Results} \label{sec:results}
\subsection{Evolution of L-maps During the Eruption}\label{sec:evol_lmaps}

\begin{figure*}[!htb]
%fig_bzmap fig_lmaps
\centering % % Answer: [trim={left bottom right top},clip]
\includegraphics[width=6.65cm,trim={0.4cm 0cm 2.0cm 1.0cm},clip]{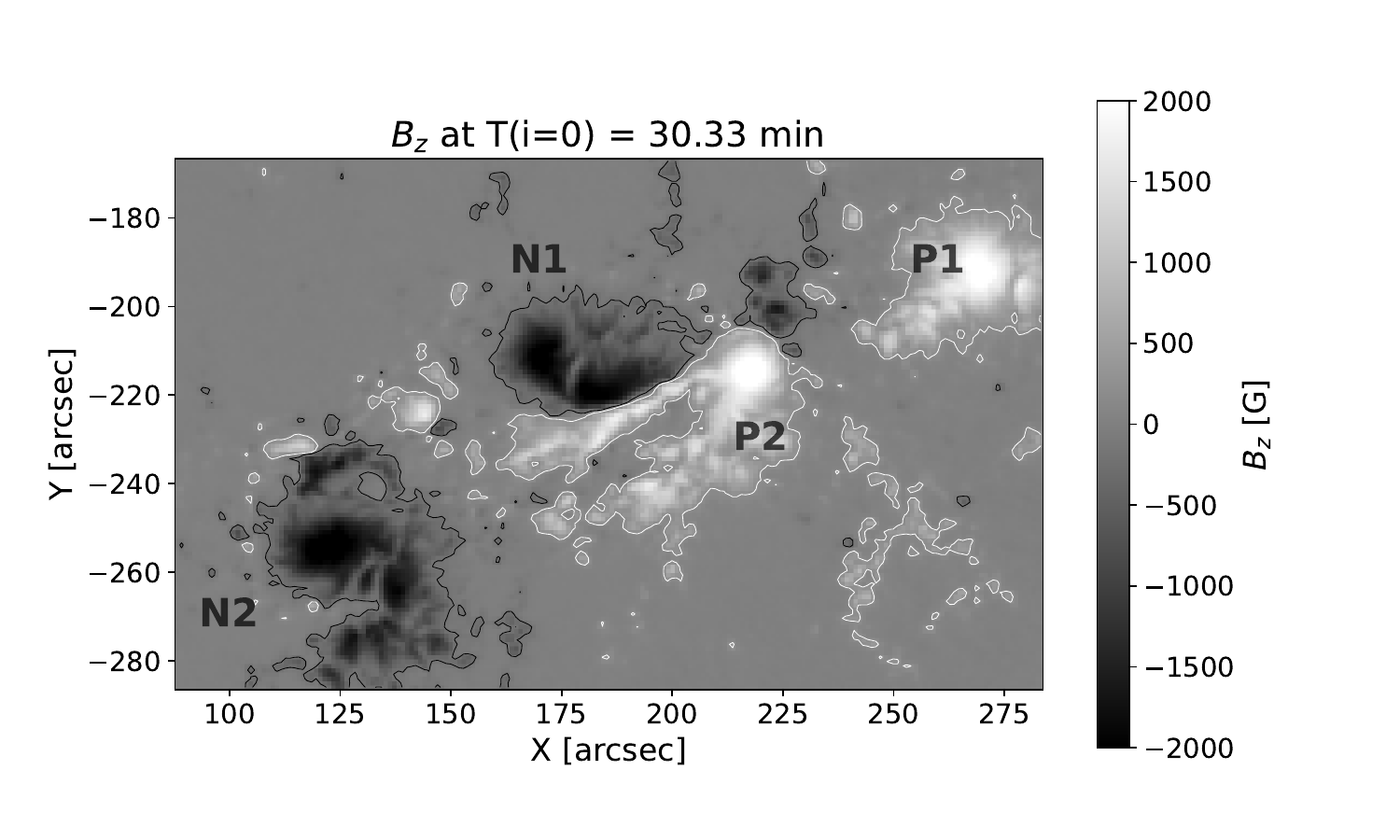}
\includegraphics[width=4.65cm,trim={3.0cm 0cm 6.3cm 1.0cm},clip]
{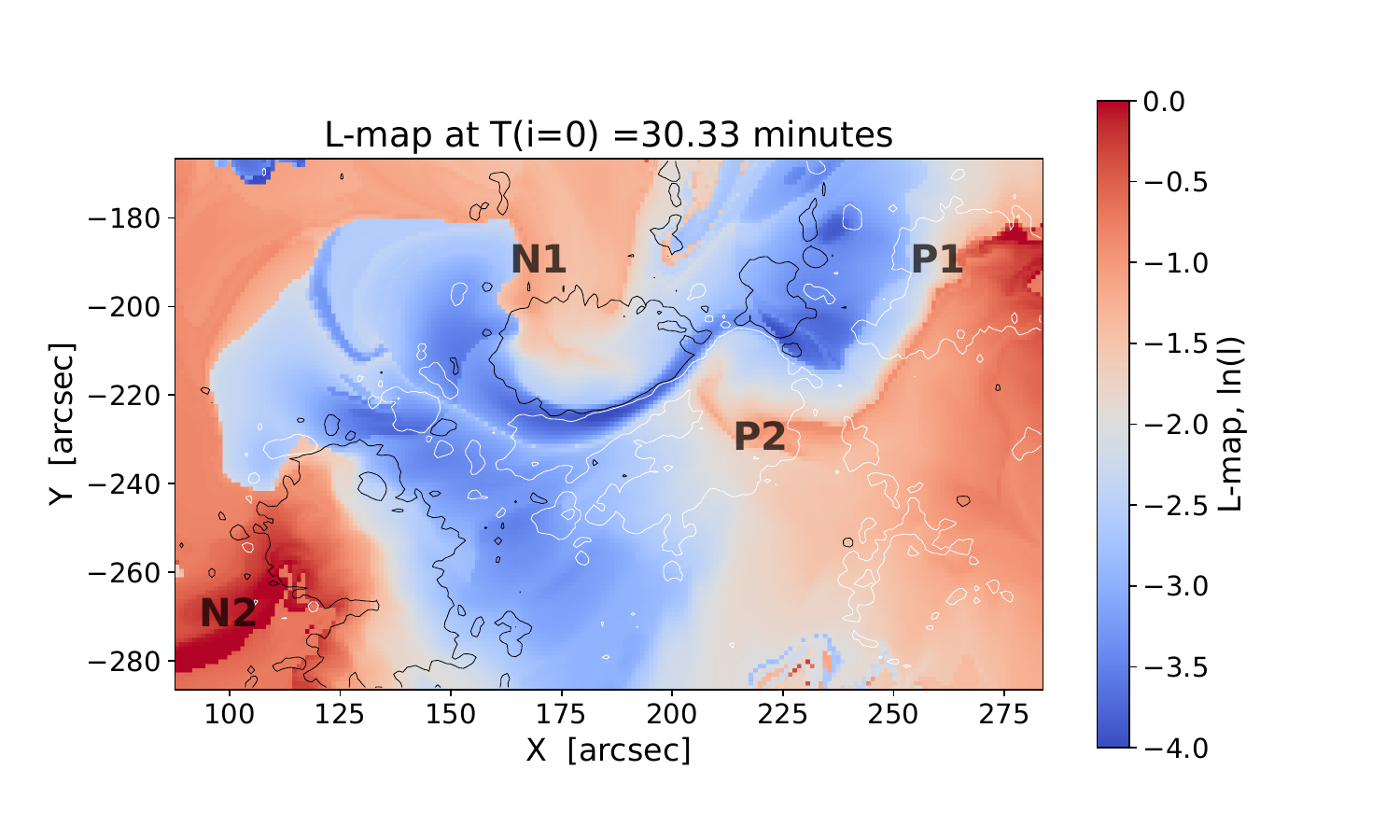}
\includegraphics[width=5.88 cm,trim={3.0cm 0cm 2.0cm 1.0cm},clip]
{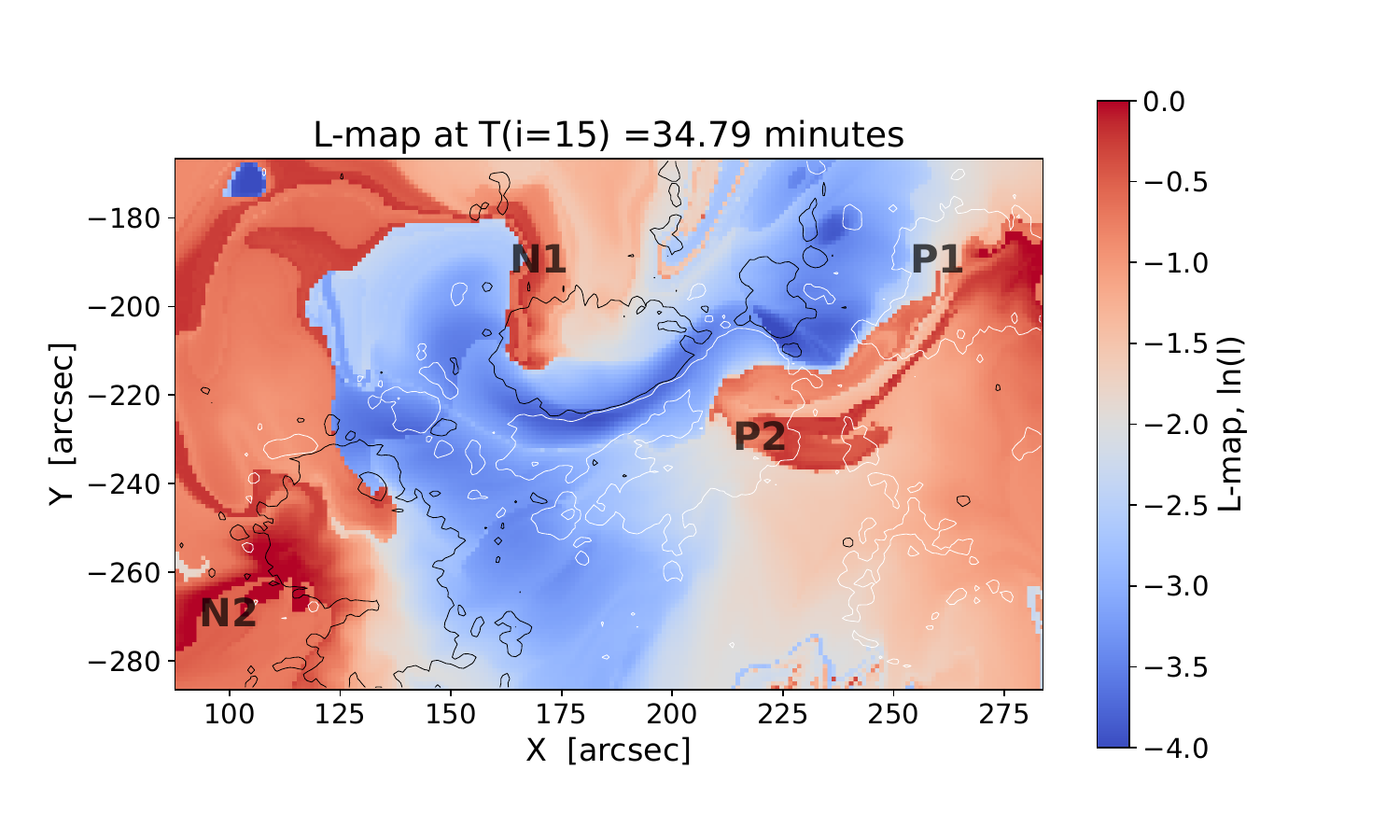}
\caption{{\it Left:} Vertical magnetic field of NOAA~AR~11158 from SDO/HMI. {\it Middle and Right:} Examples of simulated L-maps (natural logarithm of field-line length in solar radius, $\mathrm{\ln}(l)$) during the pre-flare (middle) and post-reconnection { CME expansion} (right) phases. Colors show the natural logarithm of { magnetic field-line length normalized to the solar radius}. Red corresponds to the longest field lines, while dark blue indicates the shortest ones. White and black contours mark $\pm500$~G levels of the SDO/HMI vertical magnetic field. See \S~\ref{sec:evol_lmaps} for details.} 
\label{fig:lmaps}
\end{figure*} 

Figure~\ref{fig:lmaps} presents two examples of L-maps during the pre-flare rise phase and during the post-reconnection { CME expansion} phase (see \S~\ref{sec:sim1} and \S~\ref{sec:stages}). { We find that over entire eruption L-maps, or maps of natural logarithm of magnetic field-line lengths $l$, range from $L_\mathrm{min}=\ln(l)_\mathrm{min}=-4.3$ to $L_\mathrm{max}=\ln(l)_\mathrm{max}=2.15$. These correspond to physical field-line lengths of $l_\mathrm{min}=0.01R_\odot$, $l_\mathrm{max}=8.6R_\odot$ with a mean value of field-line length $\bar{l}=0.2R_\odot$}. As the eruption progresses, the central blue region—representing short field lines between the opposite magnetic polarities N1 and P2—broadens, indicating magnetic reconnection and field-line shortening corresponding to flare ribbons. Subsequently, other portions of the blue regions near P2 and N2 become more red, reflecting field-line lengthening within the footpoints of the erupting structure, consistent with the development of coronal dimmings.

%%%%%%%%%%%%%%%%%%%%%%%%%%%%%%%%%%%%%%%%%%%%%%%%%%%%%%%%%%%%%%%%%%%%%%$
%====================RIBBONS=and DIMMINGS: OBS VS SIMULATIONS=========
\subsection{Validating Coronal Magnetic-Field Evolution Through Comparison of Observed and Simulated Flare Ribbons and Coronal Dimmings}\label{sec:val}

%-------
\begin{figure*}[!htb]
%fig03 dL_thr=0.4 and lmax2show=-0.1
%input: data_cube = np.log(lmap_cube[:,100:220,50:250])
%program: show_cum_mask
%
%TBD add Captions Ribbons: Simulation; Ribbons: Observations and similar dimming captions above each panel;
% Answer: [trim={left bottom right top},clip]
\centering 
%fig_rbn_sim_evol - need to adjust keywords
\includegraphics[width=8.8cm,trim={6.2cm 0 6.0cm 1.0cm},clip]
{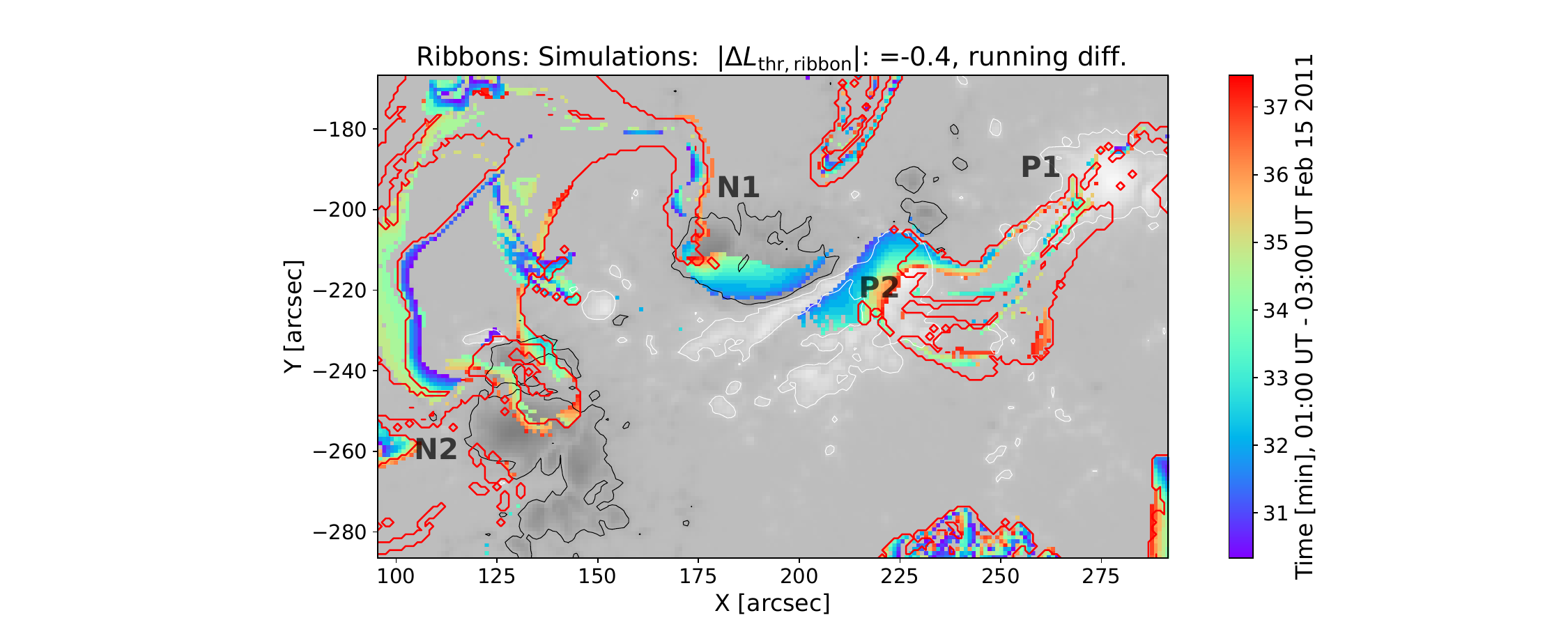}
%fig_rbn_obs
\includegraphics[width=8.6cm,trim={7.14cm 0 5.8cm 1.0cm},clip]{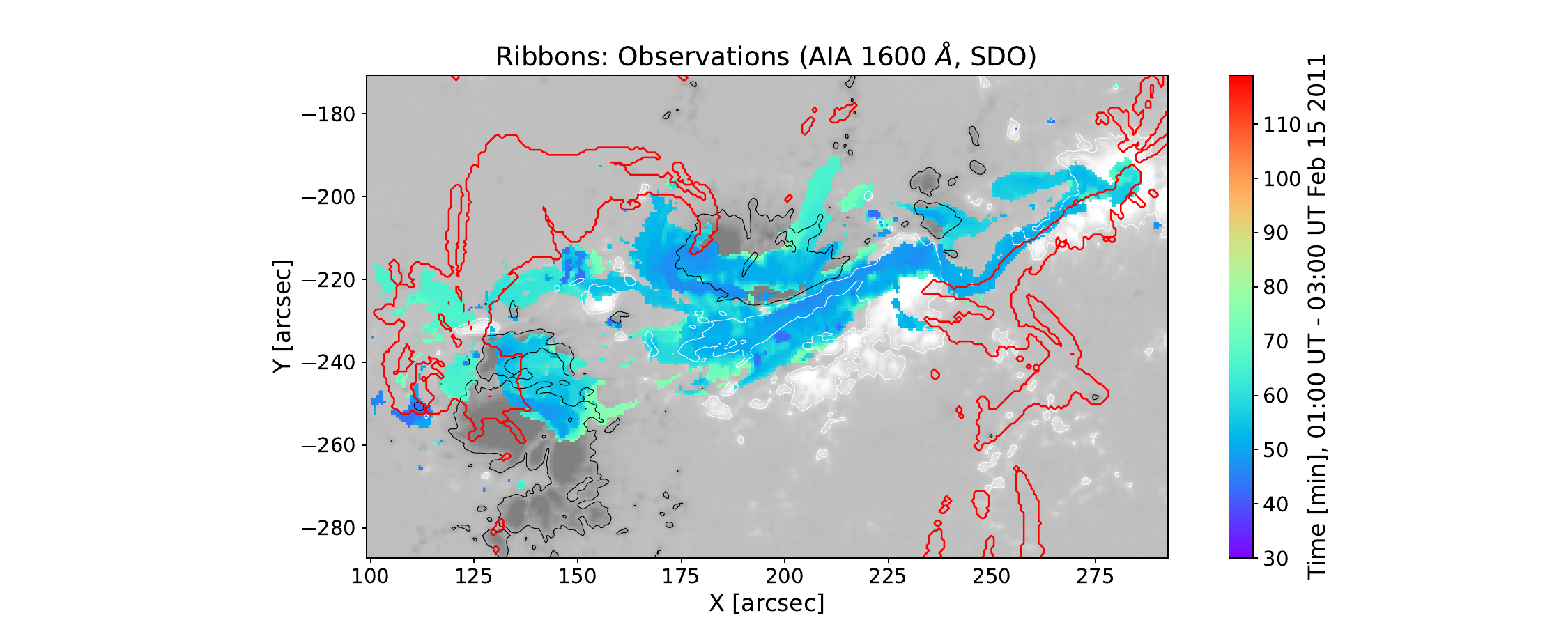}
\centering 
%fig_dim_sim_evol - need to adjust keywords
\includegraphics[width=8.8cm,trim={6.2cm 0 6.0cm 1.0cm},clip]
{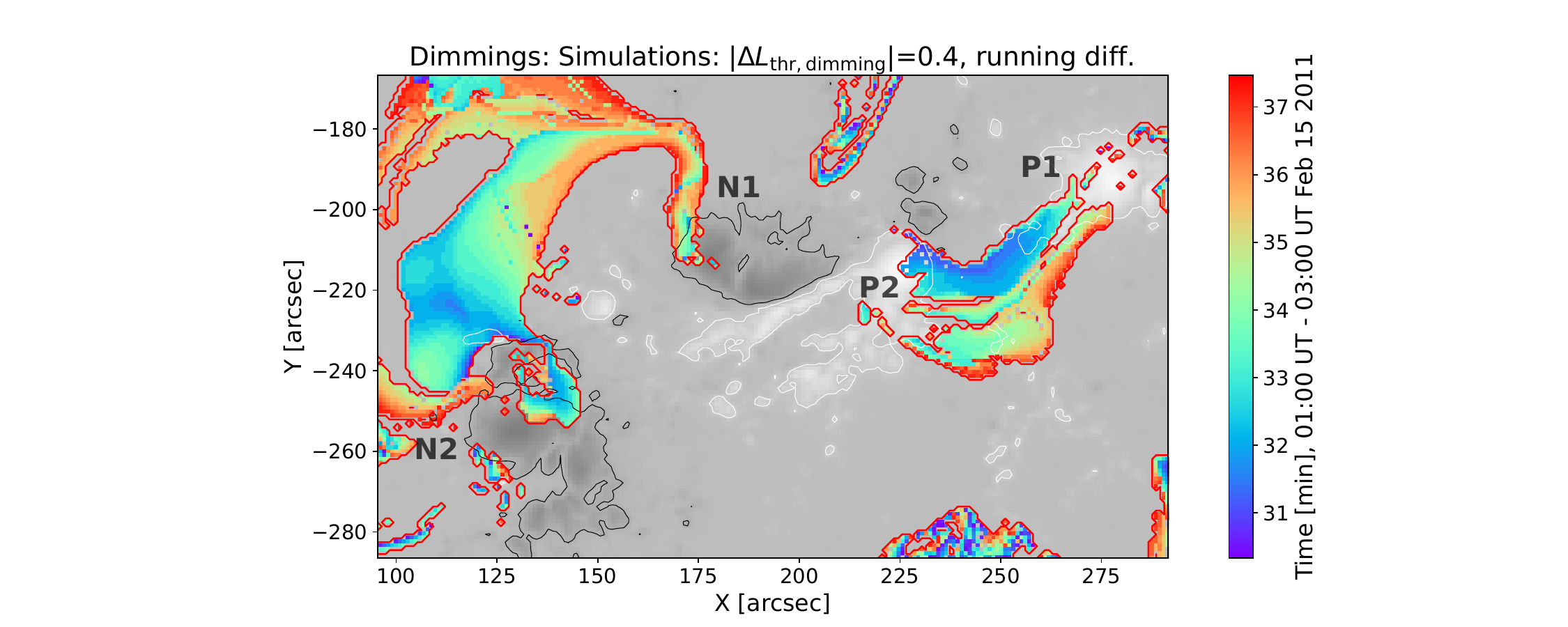}
%fig_rbn_obs_evol obs_dimming_cum_mask_evolution_time_211_log_diff.pdf 
\includegraphics[width=8.6cm,trim={7.14cm 0 5.8cm 1.0cm},clip]
{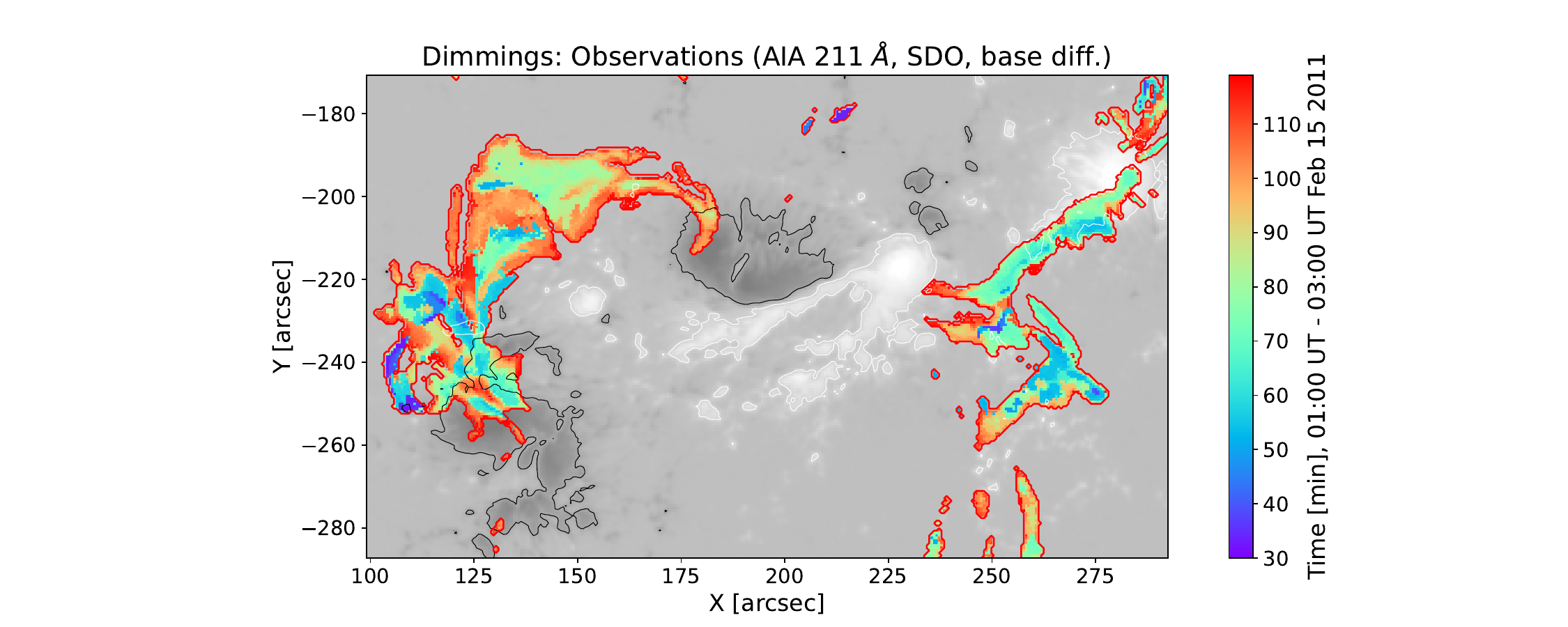}
%{obs_dim_lin_aiathr_200_diffthr_-100_ind1_1_ind2_36_ave1st_1_aventh_30.pdf}
%\includegraphics[width=8.8cm,trim={0cm 0 0.0cm 0.0cm},clip]{sim_rbn_dmn_fluxes.pdf}
%\includegraphics[width=8.8cm,trim={0cm 0 0.0cm 0.0cm},clip]{obs_rbn_dmn_fluxes.pdf}
%old one {obs_dimming_cum_mask_evolution_time.pdf}
%clusters
%\raggedright  
%\includegraphics[width=16.3cm,trim={0.0cm 0.2cm 0.5cm 0.2cm},clip]{logL_clusters_asis99.pdf}
\caption{Comparison of simulated ({\it left column}) and observed ({\it right column}) cumulative evolution masks for flare-ribbons ({\it top row}) and coronal-dimmings ({\it bottom row}). The color corresponds to the first appearance of ribbon or dimming, with blue and red colors corresponding to earlier and later times, respectively. The white and black contours show $\pm500$G-level vertical magnetic field from the HMI/SDO, respectively. The red contours outline locations of simulated ({\it left column}) and observed ({\it { right} column}) coronal dimmings. { To identify ribbons and dimmings from L-maps we used a threshold of $\Delta L_{thr, ribbon} = -0.4$ and $\Delta L_{thr, dimming} = 0.4$, respectively.} For details see \S~\ref{sec:val}.}
\label{fig:rbn_dmn}
\end{figure*} 

In Figure~\ref{fig:rbn_dmn} we compare the morphologies of flare ribbons and coronal dimmings derived from L-map simulations (left) with those identified in observations (right). To determine instantaneous ribbon and dimming locations in the simulations, we used the running difference between consecutive L-maps { with the L-map threshold of 0.4 as described in \S~\ref{sec:sim1}. To identify ribbons, we looked for field lines, where $\Delta L(t)=\ln\frac{l(t)}{l(t-\Delta t)}<\Delta L_\mathrm{thr,ribbons}=-0.4$, i.e., where field line length decreased by $e^{-0.4}=0.7$ times the initial length (see \S~\ref{sec:sim1}). To identify dimmings we looked for field lines, where $\ln\frac{l(t)}{l(t-\Delta t)}>\Delta L_\mathrm{thr,dimming}=0.4$, i.e., where field line length increased by $e^{0.4}=1.5$ times the initial length (see \S~\ref{sec:sim1}).  
From comparison in Figure~\ref{fig:rbn_dmn} we find} that the simulations reproduce the overall observed morphology and dynamics of ribbons and dimmings. Specifically, we find that:
\begin{itemize}
    \item initial two-ribbon brightenings appear close to the flare polarity inversion line (PIL), followed by gradual perpendicular motion away from the PIL; { gradual ribbon expansion in the simulation is more apparent than in the observations due to saturation, which obstructs precise ribbon identification during the flare impulsive phase.;}
    \item on the negative-polarity side, dimmings first appear near the N2 polarity and later expand toward the central N1 polarity region (transition from dark blue to red);
    \item on the positive-polarity side, dimmings develop between the P1 and P2 polarities and extend toward the south; 
    \item in both simulations and observations, ribbon brightenings and { early} dimming formation  occur nearly simultaneously { (blue and green)}, followed by gradual outward motion of dimming regions { (red)}.
\end{itemize}
%From the observations/simulation comparison we conclude that the simulations reproduce the overall observed dimming and ribbon morphology and dynamics including: (1) initial two-ribbon brightening close to the flare PIL, followed by a gradual perpendicular motion away from the PIL (transition from dark blue to lighter green);  (2) for negative polarity, initial dimming appearance close to the south-east polarity, followed by expansion to the closer center negative polarity (transition from dark blue to red); for positive polarity, the dimming appears between the two polarities on the East, followed by expansion to the central polarity (3) evolution sequence of first co-temporal ribbon brightening and dimming appearance, followed by the gradual dimming motion. 
We also find some differences between ribbons and dimming morphology in the simulations and observations including: (1) less elongated ribbons and lack of ribbon motion parallel to the PIL in the simulations compared with observations and (2) absence of early dark blue dimmings co-temporal with early ribbons in the observations, which we discuss below. { We further note that the simulated ribbon and dimming evolution occurs on a much shorter timescale ($\approx 7$~min) than observed ($\approx 60$~min). This known limitation of such simulations arises from their finite numerical resolution, which causes reconnection to proceed much faster and therefore produces ribbon evolution more than an order of magnitude quicker than in the observations. } %We discuss these differences in \S~\ref{sec:discussion}.
%%%%%%%%%%%%%%%%%%%%%%%%%%%%%%%%%%%%%%%%%%%%%%%%%%%%%%%%%%%%%%%%%
\subsection{Evolution of Field-Line-Length Clusters: { Three Stages of Flare Coronal Magnetic Field Evolution}}\label{sec:stages}

\begin{figure*}[!htb]
\centering % % Answer: [trim={left bottom right top},clip]
\includegraphics[width=18.0cm,trim={2cm 0.2cm 0cm 0.2cm},clip]{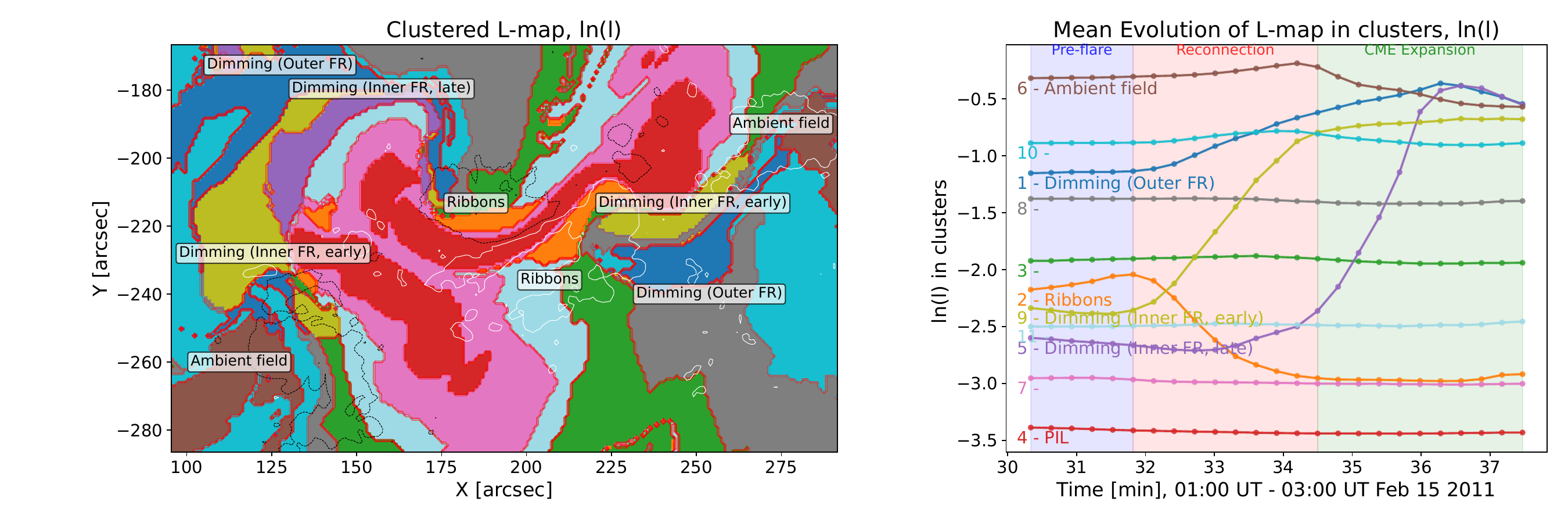}
\caption{{\it K-means clustered L-map and three stages of flare coronal magnetic-field { evolution}:} Groups of similar evolution of coronal magnetic field lines length. {\it Left:} L-map clusters. The labels mark locations of characteristic features in the 3D magnetic field evolution. {\it Right:} Mean { values of L-map} $L=\mathrm{\ln}(l)$ within individual clusters. Colors correspond to the left panel. For details see \S~\ref{sec:stages}.}
\label{fig:clusters}
\end{figure*}

In Figure~\ref{fig:clusters}(left), we show clusters of typical magnetic field line-length evolution derived from L-maps. To identify clusters, we applied the K-means clustering method \citep{MacQueen1967,Lloyd1982}. We selected 11 clusters ($n_\mathrm{clusters}=11$) as the optimal number for partitioning:  the inertia curve exhibits a clear elbow, while the silhouette coefficient \citep{Rousseeuw1987} reaches a plateau of about 0.37, indicating stable and well-separated cluster structure. The Calinski–Harabasz index \citep{CalinskiHarabasz1974} also reaches a high value of approximately 3200, further supporting this choice. Together, these clustering diagnostics indicate that 11 clusters provide a balanced and physically meaningful separation of field-line populations into ribbon and dimming regions.
%
%In Figure~\ref{fig:clusters}(left) we show clusters of typical magnetic field line-length evolution derived from L-maps. To identify clusters, we applied the K-means clustering method \citep{MacQueen1967,Lloyd1982}. We found an optimal separation of the field lines into ribbon and dimming regions using $n_\mathrm{clusters}=11$ clusters. 
%
We find that this clustering approach reveals several characteristic flare-evolution features: 
%1
low-lying field lines near the PIL (red, cluster 4), 
%2
shortening field lines within ribbons (orange, cluster 2), 
%3
and lengthening dimming field lines associated with inner and outer flux ropes (yellow and blue, clusters 9 and 1). 

Figure~\ref{fig:clusters}(right) shows the mean L-map value (logarithm of the field line length) within each cluster. By examining the evolution of the flare-ribbon cluster (orange), we identify three stages of flare coronal magnetic field dynamics { which we highlight in the accompanying 3D coronal magnetic-field image sequence (Figure~\ref{fig:3dcorona})}:
%\subsubsection{Four stages of magnetic field evolution in eruptive flare}\label{sec:stages}
%We use the map of clusters of similar coronal magnetic field evolution shown in Figure~\ref{fig:clusters} to identify the following four stages of flare coronal magnetic field dynamics:
%
\begin{itemize}
    \item{{ Slow pre-flare rise phase} ($30.3-31.8$) min: The field lines at the ribbon footpoints (orange, cluster 2) gradually stretch (Figure~\ref{fig:3dcorona}a)}. 
    \item{{ Flare reconnection with CME rise} ($31.8-34.5$) min: Reconnection causes ribbon field lines (orange, cluster 2) to suddenly shorten, while dimming field lines (yellow, cluster 9) abruptly lengthen. Outer flux rope lines (dark blue, cluster 1) expand gradually (Figures~\ref{fig:3dcorona}b,c).}
    \item{{ Post-reconnection CME expansion} ($34.5-37.3$) min: Primary flare reconnection ceases (orange, cluster 2), but inner and outer flux ropes continue to rise (clusters 5, 1, 9) (Figures~\ref{fig:3dcorona}d,e).}
    %\item{{\it Post-CME relaxation ($1.61-1.62$) hrs}: Field lines lengths are mostly constant. Ribbons field lines stretch slightly (2). Outer FR implodes (5).}
\end{itemize}
%%%%%%%%%%%%%%%%%%%%%%%%%%%%%%%%%%%%%%%%%%%%%%%%%%%%%%%%%%%%%%%%%

%-----cluster image---------
%TBD add shaded color to indicate four stages of solar flare;
% see fig04_plot and  plot_cluster_results in ribbons_lib.py to modify text

%TBD figure out the best method to identify stages in L-map evolution.

\subsection{Evolution of Individual Field Lines Anchored in Flare-Ribbons and Inner and Outer Flux-Rope Areas}\label{Lpix}

We now examine the evolution of field line lengths within individual clusters corresponding to ribbons and flux-rope dimmings across the three flare stages.

\begin{figure*}[!htb]
\centering % % Answer: [trim={left bottom right top},clip]
\includegraphics[width=15.7cm,trim={0.1cm 1.3cm 0.4cm 0.2cm},clip]{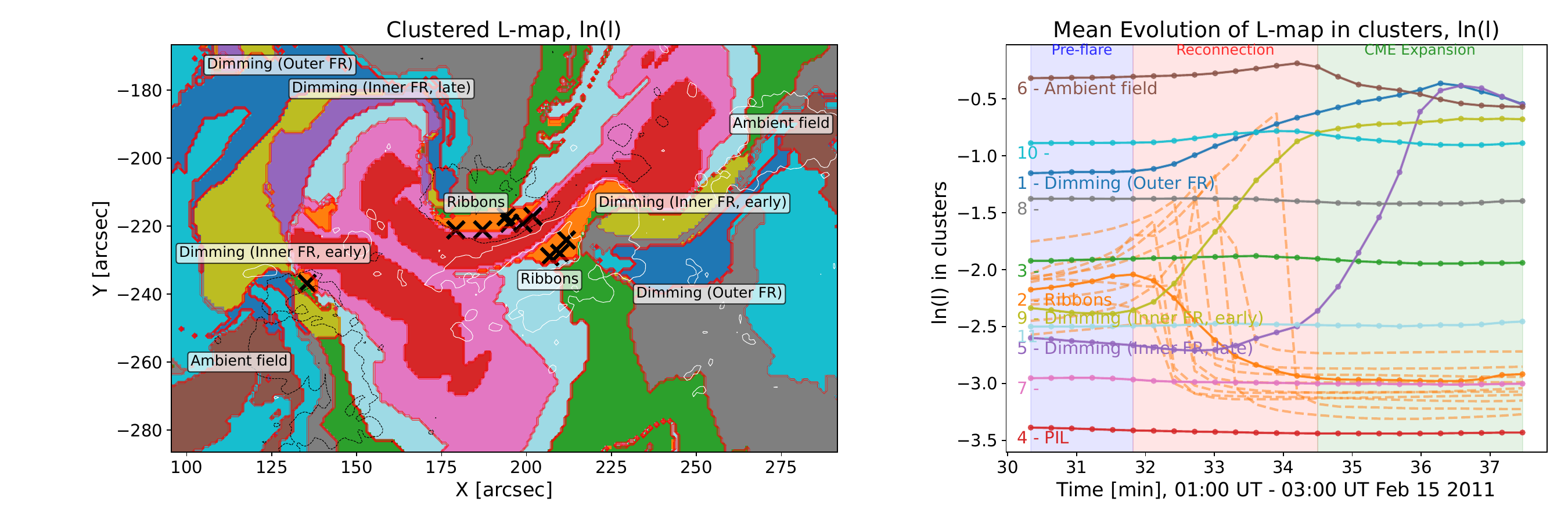}
\includegraphics[width=15.7cm,trim={0.1cm 1.3cm 0.4cm 1.2cm},clip]{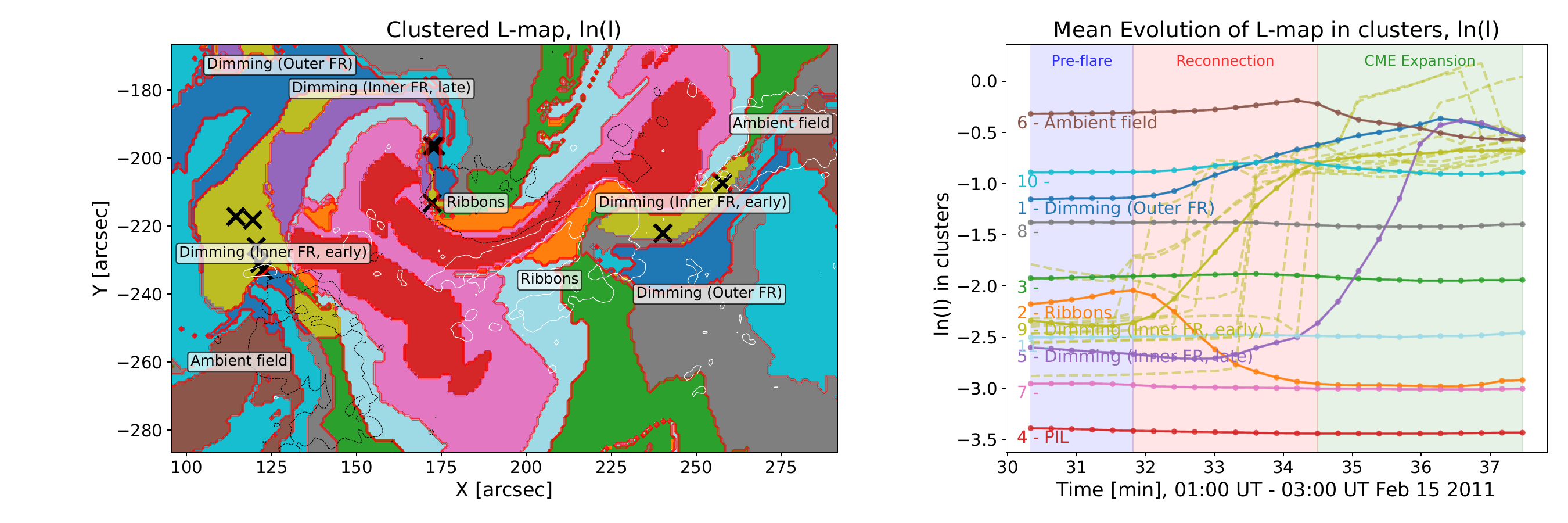}
\includegraphics[width=15.7cm,trim={0.1cm 1.3cm 0.4cm 1.2cm},clip]{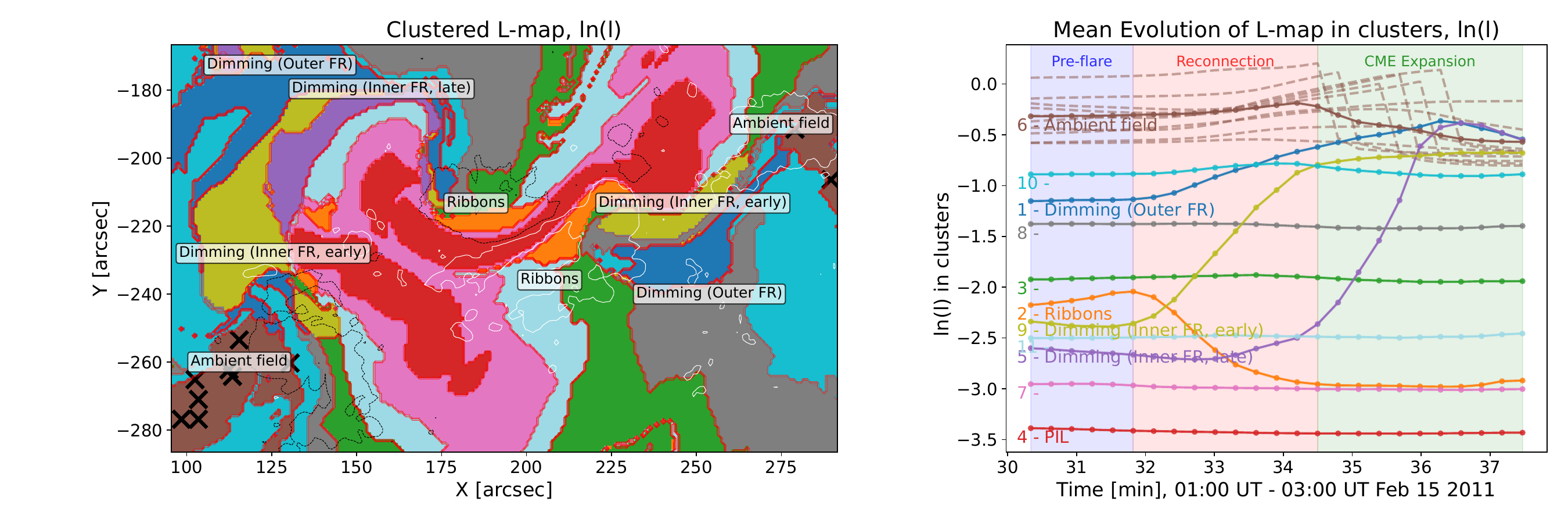}
\includegraphics[width=15.7cm,trim={0.1cm 1.3cm 0.4cm 1.2cm},clip]{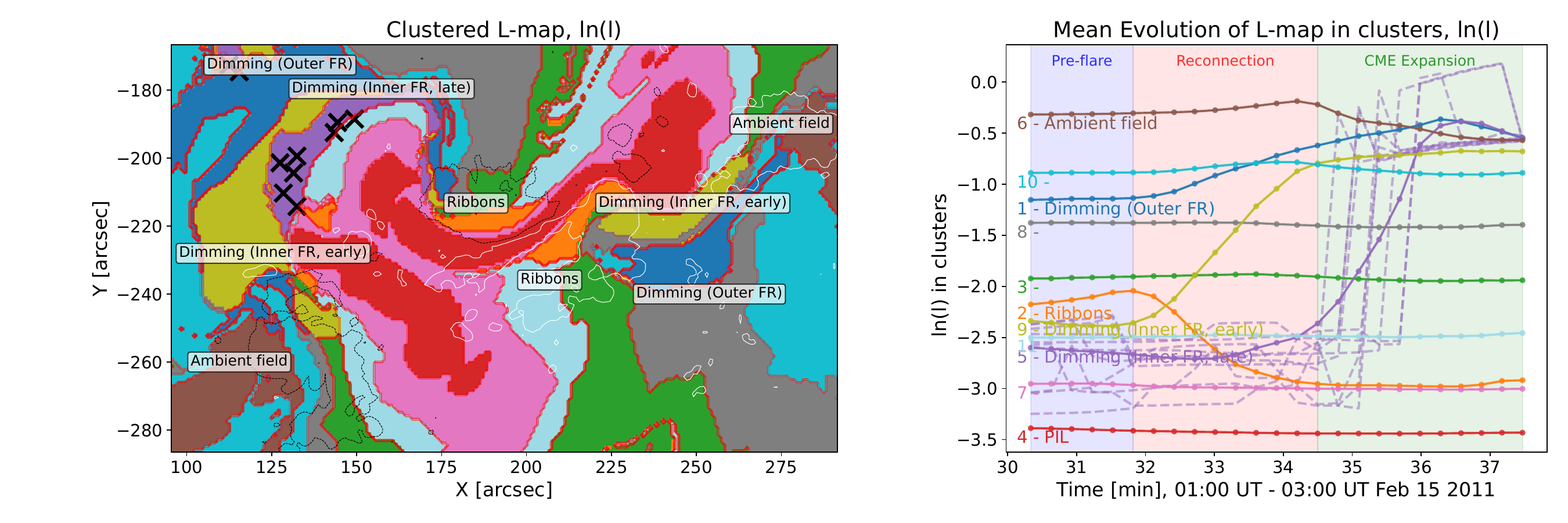}
\includegraphics[width=15.7cm,trim={0.1cm 0.2cm 0.4cm 1.2cm},clip]{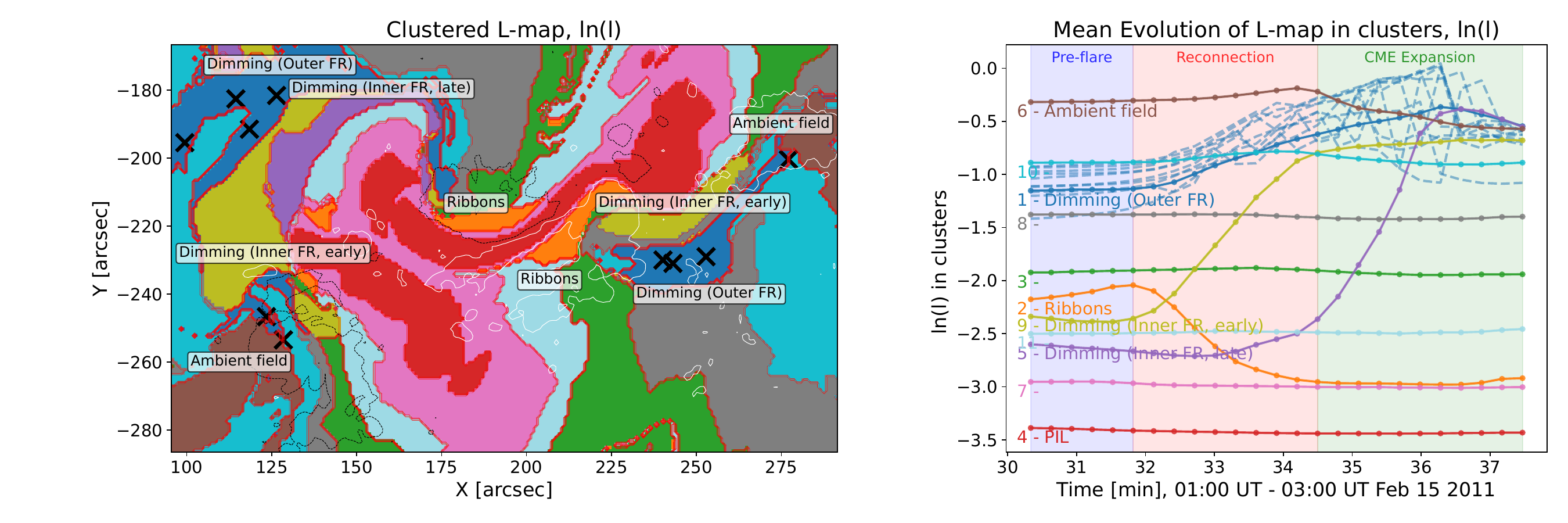}
\caption{
%Evolution of magnetic field line lengths, $L=\mathrm{\ln}(l)$, anchored in flare ribbons (first row), early inner flux rope (FR) reconnection dimming (second row), overlying large-scale ribbons (third row), inner FR reconnection dimming (fourth row) and outer FR gradual dimmings (fifth row). Note the pre-flare rise phase within flare-ribbon pixels (first row) and the reconnection dimming within inner flux-rope pixels (second row). See \S~\ref{Lpix} for details.
{ Similar to Figure~\ref{fig:clusters}, K-means clustered L-map (color, {\it left column}) with $10$ randomly selected pixels (crosses) in characteristic clusters: flare ribbons (first row), early-reconnecting inner flux rope (FR; second row), overlying ambient field (third row), late-reconnecting inner FR (fourth row), and outer FR (fifth row). The {\it right column} shows mean values of L-map, $L=\mathrm{\ln}(l)$, within individual clusters (solid line) and $10$ randomly selected pixels (dashed line). Note the pre-flare field-line lengthening in flare-ribbon pixels (first row) and the field-line lengthening associated with reconnection dimming (second row). See \S~\ref{Lpix} for details.}}
\label{clusters_rbns}
\end{figure*}

The two top rows of Figure~\ref{clusters_rbns} show evolution of L-maps in individual pixels within two clusters that actively participate in the ``flare-reconnection'' stage of coronal magnetic field dynamics. The top row shows pixels from the ribbon cluster with crosses on the left panels and dashed lines  on the right panels marking { location and evolution within} individual pixels, { respectively}. From the temporal evolution of the field line lengths within the ribbon area (top, right), we find that field lines gradually increase their lengths during the pre-flare phase, a phenomenon we call the {\it slow pre-flare rise phase}. During the second, flare-reconnection phase, the ribbon-routed field lines abruptly shorten as a result of reconnection. At the same time, pixels within the inner flux rope (yellow areas, second row, black crosses) suddenly increase in length. As this length increase occurs abruptly, we suggest that the cause of this sudden increase in the field line length is magnetic reconnection. 
As magnetic reconnection between two field lines occurs, it creates a shorter reconnected field line with flare ribbons at its footpoints  and a longer reconnected field line with coronal dimmings at its footpoints. We call the region with an increase in field line length {\it the ``reconnection dimming''}. ``Reconnection dimmings'' differ from classical coronal dimmings that arise from adiabatic plasma expansion, as the CME flux rope gradually rises \citep{Cheng2016}. 
Note that relaxing the dimming-identification threshold in log-scaled AIA 211\AA{}/SDO images ($\delta I=-0.8$ DN/pix, $I_\mathrm{max}=6.5$ DN/pix) we find an early dimming co-spatial with reconnection dimming in simulations (cluster 9), although it appears much fainter than the observed dimmings in Figure~\ref{fig:rbn_dmn}.
%Note that when we relax the dimming-identification threshold applied to log base AIA 211\AA{}/SDO images using $\delta I=-0.8$ DN/pix and $I_\mathrm{max}=6.5$ DN/pix, we find an early dimming co-spatial with reconnection dimming in simulations (cluster 9). However this dimming is quite faint compared to observed dimmings in Figure~\ref{fig:rbn_dmn}.

The three bottom rows of Figure~\ref{clusters_rbns} show the evolution of field-line lengths within clusters that evolve primarily during the {\it third stage of the flare}, corresponding to CME expansion. During this stage, some of the ambient field lines (third row, brown cluster) reconnect with inner and outer flux-rope fields (fourth and fifth rows, violet and blue clusters), leading to shrinkage of these ambient field lines and expansion of the field lines of the inner and outer flux-rope clusters. The outer flux rope, shown in the fifth row, exhibits a mix of patterns: gradual expansion during reconnection and rapid decreases, likely caused by localized reconnection events. 

%, field lines shortening and lengthening (third and fourth rows) due to reconnection of the ambient field with the inner flux rope. 

%%%%%%%%%%%%%%%%%%%%%%%%%%%%%%%%%%%%%%%%%%%%%%%%%%%%%%%%%%%%%%%%%

%\subsection{Flare Ribbons: Simulations Vs. Observations}
%\subsection{Horizontal field chang and ribbon/dimming dynamics}

\section{Discussion} \label{sec:discussion}

In this paper, we introduce { maps of natural logarithm of magnetic field-line lengths} or {\it L-maps} and show how these could be used to understand complex evolution of the 3D coronal magnetic field in realistic simulations. 
We compare ribbons and coronal dimmings from L-maps with observations, finding many similarities in their morphology, and validating our simulated coronal magnetic field evolution. We then use the validated L-maps to show how K-means clustering allows one to track and interpret complex evolution of 3D coronal magnetic fields from the simulations. Specifically, we used these maps to distinguish three stages of coronal magnetic field evolution: slow pre-flare rise phase, flare reconnection accompanied by CME rise and post-reconnection CME expansion. Furthermore, { in the simulations we find a distinctive pattern of } rapid expansion of field lines due to reconnection during the flare-reconnection stage, which we call ``reconnection dimming''. 

From the evolution of L-maps, we identify the {\it slow pre-flare rise phase}, a gradual expansion of coronal magnetic field lines prior to major reconnection. We find that magnetic field lines where L-map increases before the flare are rooted in the future ribbon areas before these participate in reconnection. This process reflects the build-up of magnetic energy and the slow rise of stressed magnetic structures before the impulsive flare onset. Several observational and theoretical studies have identified similar pre-flare rise phase in the past. The {\it theoretical basis} for the pre-flare rise phase was established in early models of sheared magnetic arcades. \citet{Klimchuk1990} demonstrated that photospheric shearing motions can lift coronal field lines upward even without reconnection, increasing the height and energy of the magnetic configuration. Later MHD simulations have confirmed that shearing, twisting, or flux emergence naturally lead to gradual coronal field rise as magnetic free energy accumulates \citep[e.g.,][]{Antiochos1999, Aulanier2010}. 

{\it Observational evidence} for slow rise of pre-flare configurations has been reported in both eruptive and confined flares (e.g., \citealt{Zhang2001,Sterling2005,McCauley2015,Kliem2014,Cheng2020,Qiu2024}). For example, using multiwavelength SDO/AIA imaging, \citet{Zhang2017} detected slow coronal dimmings and loop expansion more than an hour before an X1.6 flare, interpreting it as gradual lifting of magnetic field lines surrounding a flux rope.  Similarly \citet{Zhu2024} observed expansion of precursor dimming accompanied by rising motions of loops toward the null point, removing the strapping field above before an explosive filament eruption. 
The slow-rise phase has also been described in studies of CME initiation, where overlying loops expand quasi-statically before rapid acceleration \citep[e.g.,][]{Karpen2012,Wang2019,Cheng2023,Liu2023,Xing2024,Xing2025,Liu2025}. To summarize, slow rise phase in our simulations and earlier observations and simulations works indicates that magnetic systems evolve through a long period of quasi-static energy build-up with gradual stretching of ribbon-connected field lines before the flare onset. 
%The slow rise of pre-flare configurations observed in our simulations and L-map analysis fits naturally into this framework, representing the slow build-up and stretching of ribbon-connected field lines that precedes the flare onset.

{ From the evolution of L-maps, we also { detect} a distinct abrupt {\it reconnection dimming}, which occurs simultaneously with flare ribbons (see yellow cluster $9$ in Figure~\ref{clusters_rbns}, second row).   
This phenomenon reflects the sudden expansion of field lines that have just reconnected, marking a fundamental stage in the coronal magnetic field restructuring process. While flare ribbons trace the newly reconnected, {\it shortened field lines}, the associated dimming regions represent their counterparts -- {\it footpoints of newly formed, lengthened field lines} that become part of the erupting or expanding coronal system (as shown in Figure 10 of \citealt{Fan2024}). Along with classical dimming, this dimming adds magnetic flux to the erupting flux rope in support of the ``in-situ'' reconnection-formed flux rope scenario \citep{Nindos2020kuku,Patsourakos2020}. Reconnection dimmings are therefore distinct from classical or gradual dimmings produced solely by gradual expansion of the flux rope.}

{ The reconnection dimmings that we introduce here fits naturally within the new categorization of coronal dimmings described by \citet{Veronig2025} (their Table 1), where dimmings can arise from flux-rope expansion, reconnection, or a combination of both. Our reconnection dimming is morphologically closest to the strapping–strapping fields scenario illustrated in Figure 31a of \citet{Veronig2025}. However, in our case the reconnection dimming is produced by reconnection between the core active-region fields—as shown in Figure 10 of \citet{Fan2024}—rather than by reconnection involving the  strapping fields overlying the flux rope.}

Can { one} detect reconnection dimmings in observations? 
{ Recently, \citet{Qiu2024} analyzed SDO/AIA $304$~\AA{} observations of the upper chromosphere during one C-class and one X-class eruptive flare. They identified rapid, pixel-scale dimmings that occurred immediately after impulsive brightenings. These sequences were interpreted as signatures of reconnection with overlying arcades, where the brightenings mark energy deposition at the footpoints and the subsequent rapid dimmings indicate the opening of the overlying field lines. Here,} when we relax the dimming-identification threshold, we find an early dimming co-spatial with reconnection dimming (cluster 9) and co-temporal with flare ribbons. However, this dimming is faint compared to { gradual}  dimmings identified in Figure~\ref{fig:rbn_dmn}. Furthermore, the central part of the $211$\AA{} observations, used for dimming identification, was saturated, preventing us from analysis of the entire FOV. { We also looked for a signature of pre-dimming brightening that would observationally confirm the reconnection nature of dimming described above. However, due to saturation of the $304$~\AA{} observations, we were unable to identify such brightenings.}
These limitations highlight the importance of future saturation-free coronal observations for reliable identification of reconnection dimmings.
%MUSE?

%While in this study Future observational work 

% K-means clustering effectively separates ribbons and flux ropes, enabling quantitative simulation-observation comparison.  Discrepancies likely stem from simulation resolution and uncertainties in electric-field driving.

\section{Conclusions} \label{sec:conclusion}

Realistic three-dimensional simulations of solar active regions have revealed the urgent need for robust tools that can quantitatively and qualitatively compare complex simulation outputs with observations. In this study, we used output from the data-driven MHD simulation of the flaring active region NOAA 11158 by \citet{Fan2024} to demonstrate how maps of the natural logarithm of coronal magnetic field-line lengths—{\it L-maps}—can track the spatio-temporal evolution of flare-related magnetic fields. Our main findings are as follows.
\begin{itemize}
\item Regions where L-maps decrease over time act as effective proxies for flare ribbons, capturing the shortening of field lines caused by magnetic reconnection.
\item Regions where L-maps increase over time trace coronal dimmings, representing field-line lengthening that occurs both on short and long timescales due to magnetic reconnection and subsequent CME expansion, respectively.
\item Comparison between simulated ribbons and dimmings derived from simulation L-maps and observations shows overall morphological and dynamic agreement, validating the realism of the simulated coronal evolution.
\item Applying clustering analysis to L-maps, we identify three key stages of coronal magnetic-field evolution during flares: (1) a slow pre-flare rise phase, (2) flare reconnection accompanied by CME acceleration, and (3) post-reconnection CME expansion.
\item We find clear evidence for a {\it slow pre-flare rise phase}, marked by gradual field-line stretching within ribbon footpoints prior to reconnection, consistent with prior modeling and observations.
\item We also identify {\it reconnection dimming} occurring during the flare-reconnection phase as field lines lengthen abruptly { as the result of the} reconnection { within the active region core}. Further observational studies, particularly of unsaturated coronal emission, are needed to characterize this new signature. 
\end{itemize}

We conclude that L-maps provide a powerful and physically intuitive framework for bridging simulations and observations, enabling quantitative tracking of the three-dimensional evolution of coronal magnetic fields during solar flares.

\begin{acknowledgments}
\section*{Acknowledgments}
M.K.D. acknowledges support from NASA ECIP NNH18ZDA001N and NSF CAREER SPVKK1RC2MZ3  awards. Y.F.'s work is supported by the National Center for Atmospheric Research (NCAR), which is a major facility sponsored by the U.S. National Science Foundation (NSF) under Cooperative Agreement No. 1852977. A.N.A. acknowledges support from NASA ROSES Early Career Investigator Program (grant No. 80NSSC12K0460), and the DKIST Ambassadors Program, in which funding is provided by the NSO, a facility of the National Science Foundation, operated under Cooperative Support Agreement number AST-1400450.

\end{acknowledgments}

\bibliographystyle{aasjournal}
%\bibliography{main}

\bibliography{main}

%\bibliography{bibliography,full_lib_thes_cp}{}
%\bibliographystyle{aasjournal}

\end{document}